\documentclass[a4paper]{article}
\usepackage{xcolor}
\usepackage[utf8]{inputenc}
\usepackage[T1]{polski}
\usepackage{graphicx}\graphicspath{{img/}}
\usepackage{subcaption}
\usepackage{lmodern,fancyhdr,secdot,float,amsmath,amsfonts,amsthm,amssymb}
\usepackage{booktabs,tabularx}
\usepackage[margin=2cm]{geometry}
\usepackage{multicol}
\usepackage[hidelinks,unicode]{hyperref}
\usepackage{xurl}
\usepackage{titlesec}
\titleformat{\section}{\centering\normalsize\bf}{\thesection.}{.5em}{\MakeUppercase}
\titleformat*{\subsection}{\bf\normalsize\selectfont}
\titleformat*{\subsubsection}{\bf\normalsize\selectfont}
\newcommand{\titlePL}[1]{\large\textbf{ #1}}
\newcommand{\titleEN}[1]{\normalsize #1}
\newcommand{\keywordsPL}[1]{\small\textbf{Słowa kluczowe:} #1}
\newcommand{\keywordsEN}[1]{\small\textbf{Keywords:} #1}
\newcommand{\abstractPL}[1]{\small\textbf{Streszczenie:} #1}
\newcommand{\abstractEN}[1]{\small\textbf{Abstract:} #1}

\definecolor{logo_color}{RGB}{40, 69, 166}

\begin{document}\thispagestyle{empty}\pagestyle{fancy}
\begin{minipage}[t]{0.5\textwidth}\vspace{0pt}%
\includegraphics[scale=0.9]{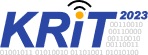}
\end{minipage}
\begin{minipage}[t]{0.45\textwidth}\vspace{12pt}%
\centering
\color{logo_color} KONFERENCJA RADIOKOMUNIKACJI\\ I TELEINFORMATYKI\\ KRiT 2024
\end{minipage}

\vspace{1cm}

\begin{center}
\titlePL{Efektywne energetycznie wielodostępowe przetwarzanie brzegowe w sieci 5G}

\titleEN{Energy efficient Multi-access Edge Computing in 5G network}\medskip

Paweł Kryszkiewicz$^{1}$

\medskip

\begin{minipage}[t]{0.7\textwidth}
\small $^{1}$ Politechnika Poznańska, Poznań, \href{mailto:email}{pawel.kryszkiewicz@put.poznan.pl}
\end{minipage}

\medskip

\end{center}

\medskip

\begin{multicols}{2}
\noindent
\abstractPL{
Wielodostępowe przetwarzanie brzegowe jest techniką łączącą wykorzystanie sieci komunikacyjnych i oddalonych zasobów obliczeniowych. Pozwala wykonać złożone zadania obliczeniowe na potrzeby urządzeń o niewielkiej mocy obliczeniowej przy zachowaniu niewielkich opóźnień. Istotne jest jednak efektywne zarządzanie przydziałem zadań obliczeniowych do poszczególnych węzłów. W pracy przedstawiono jak system przetwarzania brzegowego może być zintegrowany z siecią 5G, a także jak można rozdzielić zasoby między poszczególne węzły, żeby zminimalizować zużycie energii. Przedstawiony zostanie szereg nowych stopni swobody, które umożliwiają znaczne obniżenie zużycia energii w stosunku do istniejących rozwiązań niezależnej optymalizacji części obliczeniowej i komunikacyjnej.
\footnote[1]{Praca powstała w~ramach projektu OPUS finansowanego przez Narodowe Centrum Nauki nr 2021/41/B/ST7/00136.}}
\medskip

\noindent
\abstractEN{
Multi-access edge computing is a technique that combines the use of communication networks and remote computing resources. It allows to perform complex computational tasks for devices with low computing power while maintaining low latencies. However, it is important to effectively allocate the computing tasks to individual nodes. The work will present how the multi-access edge computing system can be integrated into the 5G network, as well as how resources can be distributed between individual nodes to minimize energy consumption. Some new degrees of freedom will be presented, which enable a significant reduction in energy consumption compared to existing solutions for independent optimization of the computation and communication parts.
}
\medskip

\noindent
\keywordsPL{Wielodostępowe przetwarzanie brzegowe, efektywność energetyczna, 5G}
\medskip

\noindent
\keywordsEN{Multi-access Edge Computing, energy efficiency, 5G}

\section{Wstęp}

Istotą systemów telekomunikacyjnych jest przesył informacji, ale co raz częściej ich źródłem jest urządzenie np. czujnik, a ujściem serwer dokonujący obliczeń albo magazynowania. Ten rodzaj usług jest zwyczajowo nazywany Internetem rzeczy IoT (ang. Internet of Things). Taka zmiana jest po części spowodowana zwiększeniem dostępności usług komunikacyjnych, a także obniżeniem kosztu zakupu i utrzymania urządzeń. Istotna jest też zmiana w rodzaju przetwarzania danych. Często dane pochodzące z tanich, zasilanych bateryjnie urządzeń wymagają złożonego przetwarzania z użyciem modeli uczenia maszynowego. Funkcje takie pełni przetwarzanie chmurowe (ang. cloud), które jest już powszechnie używane. Trzeb jednak pamiętać, że często chmurowe centra danych oddalone są o tysiące kilometrów od miejsca z którego wysyłane są dane źródłowe. Dodatkowo często po przetwarzaniu jego wynik np. decyzja o wykryciu danego obiektu w strumieniu wideo, musi być użyta lokalnie do podjęcia pewnych akcji. Taka pętla decyzyjna może powodować duże opóźnienia, a także zmienność opóźnienia, co jest nieakceptowalne dla wielu usług, w tym czasu rzeczywistego.   

\begin{figure}[H]
\centering
\includegraphics[width=3.0in]{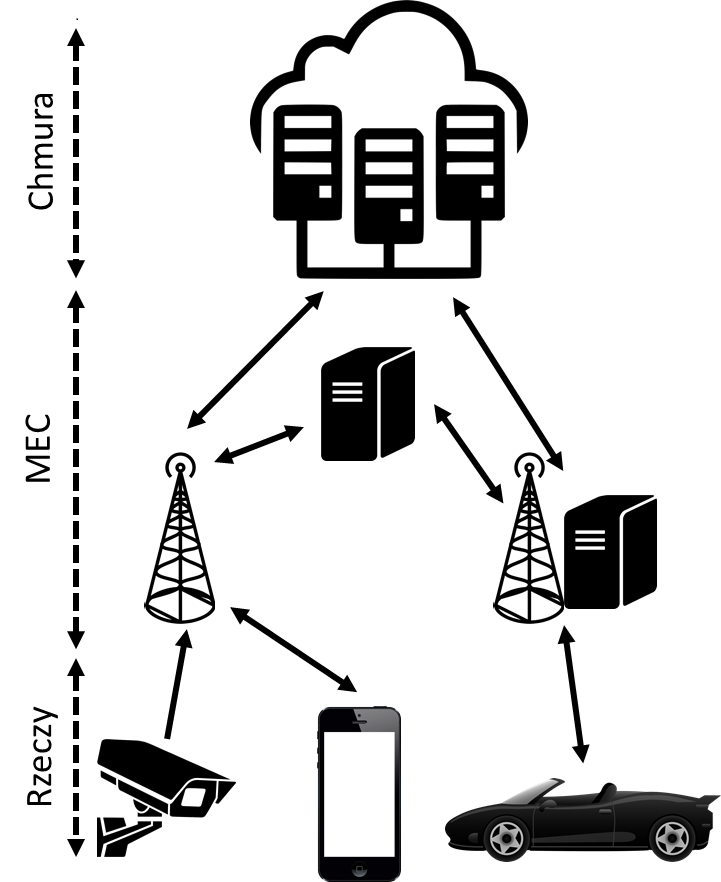}
\caption{ Warstwowa architektura brzegowej sieci obliczeniowej MEC (ang. Multiple Access Edge Computing) i chmury.}
\label{fig:MEC_tier}
\end{figure}

Żeby sprostać tym wymaganiom zaczęto rozważać wykorzystanie niewielkich centrów danych albo pojedynczych komputerów położonych bliżej urządzeń końcowych, które można nazwać "Rzeczami" poprzez odniesienie do "Internetu rzeczy". Taki układ sieci zaprezentowano na Rys. \ref{fig:MEC_tier}. Z uwagi na położenie tej warstwy pomiędzy "Rzeczami" oraz chmurą obliczeniową część badaczy odnosi się często do "mgły obliczeniowej" (ang. Fog computing)\cite{Vaquero_fog_2014}. Jednocześnie podnoszone jest, że dodatkowe węzły obliczeniowe muszą się znaleźć blisko brzegu sieci dostępowej co spowodowało nazwanie tej funkcji wielodostepowym przetwarzaniem brzegowym MEC (ang. Multiple-access Edge Computing)\cite{Shirazi_MEC_survey_2017}. Warto tu podkreślić, że często można znaleźć publikacje, gdzie rozróżnia się przetwarzanie brzegowe od mgły obliczeniowej z różną granicą podziału. Czasem autorzy sugerują, że mgła to rozrzucone elementy chmury obliczeniowej, a warstwa obliczeń brzegowych odbywa się na samych urządzeniach końcowych. Podnoszone jest też to, że pierwotnie nazwa fog była częściej używana w nawiązaniu do sieci lokalnych LAN (ang. Local Area Network), a MEC w odniesieniu do sieci komórkowych. Ciekawym jest też zmiana pierwotnego wyjaśnienia skrótu MEC z "Mobile Edge Computing" na "Multiple-access Edge Computing" przez ETSI w roku 2017 \cite{ETSI_MEC_change_2017}. Miało to podkreślić otwarcie się tworzonych standardów na sieci inne niż komórkowe. Te różnice nomenklaturowe nie są jednak znaczące z perspektywy tego artykułu.  

Technologia MEC znajduje wiele zastosowań \cite{hu2015mobile} od przyspieszania przetwarzania strumienia wideo, poprzez obsługę inteligentnych samochodów po technologię rozszerzonej rzeczywistości. Pojawiły się również zmiany w standardach zapewniające jej integrację w ramach architektury sieci 4G albo 5G \cite{3gpp_23501}. Z perspektywy badawczej jednym z problemów jest efektywny rozdział pojawiających się zgłoszeń obliczeniowych między różne węzły obliczeniowe. Istotna jest tu zarówno część sieciowa (opóźnienie i obciążenie poszczególnych węzłów albo łączy) jak również obliczeniowa (zużycie pojedyncze jednostki obliczeniowe). Efektywność można rozumieć na wiele sposobów. Z pewnością istotne jest zapewnienie użytkownikom usługi w zadanych ramach czasowych (ograniczenie na opóźnienie), ale może być też istotna minimalizacja użycia pewnych węzłów (np. w przypadku taryfikacji o różnym koszcie pracy węzłów obliczeniowych) albo maksymalizacja liczby obsłużonych zadań obliczeniowych\cite{Luo_2021_survey}. Jednym z istotniejszych podejść jest przydział zasobów minimalizujący, przy zachowaniu wymogów jakościowych, zużycie energii. Poza zmniejszeniem emisji dwutlenku węgla do atmosfery może to znacząco ograniczyć koszty ponoszone przez użytkowników, lub operatorów sieci. Z uwagi na połączenie optymalizacji przesyłu i obliczeń pojawia się wiele dodatkowych zmiennych, które można modyfikować zwiększając potencjalne ograniczenie w zużycia energii. 

W poniższej pracy zostanie przedstawione jak technika MEC może być zastosowana w sieciach 5G (Rozdział 2), a także na podstawie wcześniejszych badań autora, jak można przydzielić zadania obliczeniowe w systemie MEC zapewniając minimalizację zużycia energii.

\section{Wielodostępowe przetwarzanie brzegowe w systemie 5G}
\label{sec:MEC}
Choć przetwarzanie brzegowe może być zastosowane w różnych sieciach dostępowych warto tu przedstawić jego połączenie z siecią 5G. Z uwagi na tworzenie architektury 5G oraz architektury MEC prawie równolegle, choć przez różne organizacje, są one najbardziej kompatybilne. Oczywiście można by traktować węzły obliczeniowe MEC jako standardowe serwery o pewnych adresach IP. Z perspektywy sieci 5G dostęp odbywałby się jak do innych usług internetowych poprzez scentralizowane User Plane Function (UPF) znajdujące się w sieci rdzeniowej 5G. Nie umożliwiło by to jednak ograniczyć opóźnień na przetwarzanie MEC.

\begin{figure}[H]
\centering
\includegraphics[width=3.0in]{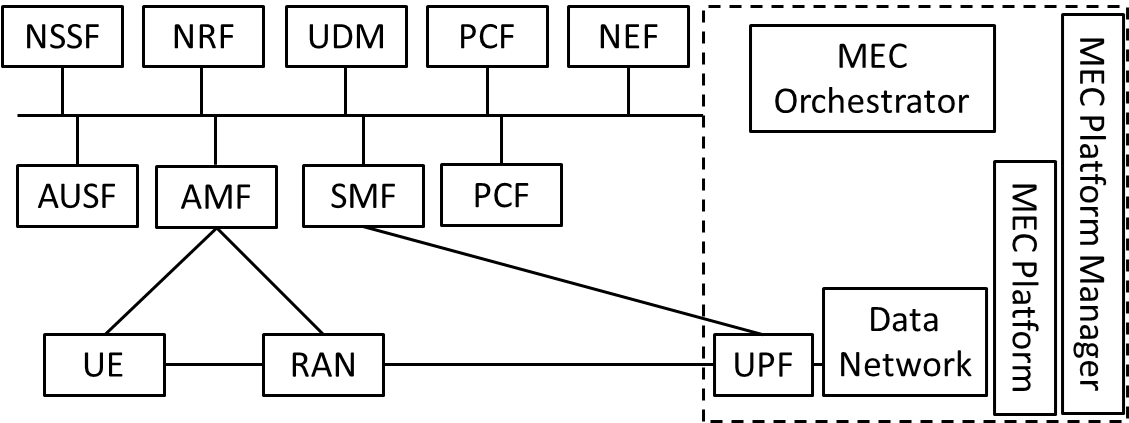}
\caption{ MEC zintegrowane z architekturą rdzeniową 5G. Na podstawie: \cite{kekki2018mec}}
\label{fig:MEC_5G_arch}
\end{figure}

Zarówno ETSI, standaryzując system MEC \cite{kekki2018mec}, jak i 3GPP standaryzując sieć 5G \cite{3gpp_23501} zapewniły jednak szereg rozwiązań poprawiających wydajność zastosowania MEc w sieci 5G. Architekturę tego połączenia przedstawiono na Rys. \ref{fig:MEC_5G_arch}. Lewa strona rysunku to standardowa architektura sieci 5G przedstawiona z uwagi na oferowane usługi. Po prawej stronie, w ramce, przedstawione są elementy składowe systemu MEC. Płaszczyzna sterująca systemu MEC tzn. MEC Orchestrator, jest połączona z płaszczyzną sterującą sieci 5G będącej szeregiem funkcji np. NEF (ang. Network Exposure Function), SMF (ang. Session Management Function). Z perspektywy sieci 5G system MEC jest podłączony jako funkcja aplikacji AF (ang. Application Function). Pozwala to systemowi MEC na dostęp do wewnętrznych funkcji systemu 5G w tym obserwować parametry radiowe np. jakość sygnału.  

Natomiast w płaszczyźnie danych zadanie obliczeniowe pojawiające się w urządzeniu końcowym UE (ang. User Equipment) jest przekazywane przez sieć dostępową RAN (ang. Radio Access Network) do odpowiedniej instancji funkcji UPF, która jest podłączona do węzłów obliczeniowych (na rysunku ang. Data network). Co istotne, może być wiele instancji funkcji UPF w całej sieci. Oznacza to, że dane UE nie muszą być fizycznie przekazane do scenatralizowanego serwera pełniącego rolę sieci rdzeniowej. Instancja UPF może być elementem wybranych stacji bazowych. Może być też zlokalizowana w przełącznikach sieciowych agregujących ruch w sieci 5G. Z uwagi na architekturę opartą na funkcjach można bardzo elastycznie dodawać węzły obliczeniowe blisko urządzeń końcowych wymagających niskiego opóźnienia usług świadczonych przez system MEC. Najbardziej oczekiwana jest instalacja węzłów MEC przez operatorów sieci komórkowych z uwagi na dostępną infrastrukturę w postaci serwerowni, stacji bazowych itp., a także możliwość dywersyfikacji oferowanych usług. Standard MEC umożliwia jednak również integrację zewnętrznych węzłów obliczeniowych. Może mieć to uzasadnienie szczególnie przy braku lokalnej infrastruktury operatora albo dla ograniczenia kosztów zarządzania węzłami obliczeniowymi. W takim przypadku system MEC nie może być traktowany jako bezpieczny z perspektywy sieci rdzeniowej 5G. Punktem dostępu staje się wtedy funkcja NEF.

Ciekawym aspektem, który postaram się rozwinąć w dalszych rozdziałach artykułu, jest rozdział zadań obliczeniowych między węzły obliczeniowe. Istniejące standardy umożliwiają takie działanie. System MEC przesyła odpowiednie polityki w warstwie sterującej dotyczące konkretnego ruchu tzn. do którego węzła UPF ma zostać przekierowany. Żeby system MEC podjął optymalną albo choć zbliżoną do optymalnej, informację o rozdziale zadań obliczeniowych między węzły musi jednak posiadać pewne informacje o warstwie dostępowej sieci radiowej. Standard przewiduje dostęp do informacji dostępnych nawet w elementach składowych sieci dostępowej tzn. CU (ang. Centralized Unit) oraz DU (ang. Distributed Unit). Jak zostanie pokazane w następnym rozdziale efektywny przydział zasobów w sieci MEC wymaga informacji np. o opóźnieniach wynikających z użycia konkretnych interfejsów radiowych chociażby poprzez ograniczoną przepływność.      

Jako że sieć 5G może obsługiwać szybko poruszających się użytkowników system MEC powinien być również dostosowany do takiej ewentualności np. zapewniając usługi MEC dla komunikacji między pojazdami. W \cite{kekki2018mec} przedyskutowano jak zapewnić ciągłość usług MEC zakładając przełączenie między węzłami obliczeniowymi. Najprostsze rozwiązanie może być zapewnione gdy dane zadanie obliczeniowe jest "bezpamięciowe" tzn. węzeł nie musi posiadać informacji o wcześniejszych przesłanych informacjach ani podjętych decyzjach. Można wtedy przenieść zadanie MEC do innej instancji MEC bez koordynacji. W przypadku zadań "pamięciowych" konieczne jest skopiowanie całego stanu danej aplikacji MEC między węzłami obliczeniowymi, a następnie jej uruchomienie. W ten sposób w sieci może pracować kilka identycznych aplikacji. W zadanym momencie jedna z aplikacji jest wyłączana zachowując ciągłość usług. W tym przypadku konieczne jest jednak stworzenie odpowiedniej aplikacji która może być powielona w sieci bez powodowania przekłamań.

\section{Modelowanie systemu MEC} 
\label{sec:system}
Przeprowadzenie optymalizacji systemu MEC wymaga zbudowania jego modelu matematycznego z perspektywy interesujących miar jakości oraz zmiennych optymalizacyjnych. Z uwagi na inną naturę przetwarzania inaczej należy również zamodelować część komunikacyjną i przetwarzania brzegowego. Najważniejszymi z perspektywy tego artykułu miarami jakości są opóźnienie czasowe oraz zużycie energii. Trzeba być również świadomym, że modele te muszą stanowić kompromis między dokładnością odwzorowania rzeczywistych urządzeń i zjawisk, a prostotą optymalizacji. 
\subsubsection{Zużycie energii na obliczenia}
\label{sec_model_energia_obliczanie}
\begin{figure}[H]
\centering
\includegraphics[width=3.0in]{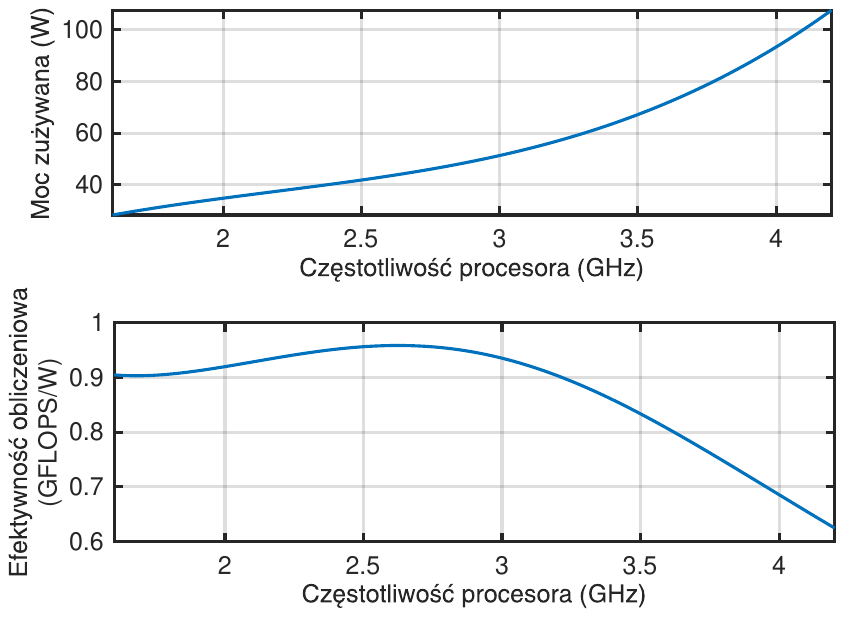}
\caption{ Moc używana i efektywność obliczeniowa dla procesora Intel i5-2500K na podstawie \cite{Kopras_TCOMM_2022}.}
\label{fig:CPU_eff}
\end{figure}

Należy przyjąć, podobnie jak w \cite{Kopras_sensors_2024,Kopras_TCOMM_2022}, że pewne zadanie obliczeniowe charakteryzuje się pewną długością $L$ bitów oraz intensywnością zadania $\theta$ wyrażoną w wymaganych operacjach zmiennoprzecinkowych FLOP (ang. Floating Point Operations) na bit. Dla danego węzła obliczeniowego można wyznaczyć efektywność obliczeniową $\beta$ wyrażoną w liczbie operacji zmiennoprzecinkowych na sekundę (często oznaczonej jako FLOPS) na Watt. Wypadkowa moc wymagana na przeprowadzenie tych obliczeń wynosi:
\begin{equation}
    P_{\mathrm{cp}}=\frac{L\theta}{\beta}.
\end{equation}
Warto podkreślić, że parametr $\beta$ zależy nie tylko od wykorzystywanej architektury obliczeniowej, ale też konkretnego algorytmu. Zgodnie z rankingiem Green 500, informującej o najbardziej wydajnych superkomputerach, aktualnie osiągane maksimum, na czerwiec 2024, to ok. 72 GFLOPS/Watt. Wydajność ta jest zależna m.in. od częstotliwości pracy procesora, a także liczby operacji wykonywanych w jednym cyklu pracy. Co istotne, w pracach \cite{Kopras_sensors_2024,Kopras_TCOMM_2022} pokazano, że można rozważyć dobór optymalnej częstotliwości pracy procesora tzn. $\beta$ jest funkcją (np. wielomianową) częstotliwości taktowania procesora. Przykładowo dla procesora i5-2500K pokazano na Rys. \ref{fig:CPU_eff} zmianę efektywności obliczeniowej i mocy zużywanej w funkcji częstotliwości pracy. 
Maksymalną efektywność procesor osiąga ok. 2.7 GHz, ale może też pracować z wyższym zegarem osiągając większą szybkość przetwarzania danych, ale przy nieproporcjonalnym wzroście zużycia mocy. 

Z perspektywy przydziału zasobów warto zwrócić uwagę, że dla zadań obliczeniowych o łagodnych wymaganiach opóźnieniowych możliwe jest przydzielenie ich do wysokoefektywnych węzłów obliczeniowych. W przypadku zadań wymagających obliczeniowo i czasowo konieczne może być użycie węzła o duże wydajności, a niższej efektywności energetycznej. Dodatkowo można wtedy też podwyższyć częstotliwość taktowania procesora. Może być to też scenariusz konieczny w przypadku dużego obciążenia sieci. Powyższy model może być z powodzeniem używany zarówno dla obliczenia w warstwie MEC jak i w chmurze uwzględniając inne parametry poszczególnych jednostek obliczeniowych. 

\subsubsection{Zużycie energii na komunikację}
Modelowanie zużycia energii przez sieć komunikacyjną jest dość złożonym problemem. W przypadku sieci bezprzewodowej stosuje się albo modelowanie każdego elementu składowego przetwarzania np. wyprowadza się wzory na zużycie energii przetworników Analogowo-Cyfrowych w danej architekturze i konfiguracji \cite{Nossek_2011} albo przeprowadza się pomiary całych urządzeń uwzględniając możliwe do zmiany parametry np. przepływność użytkowników, odległość użytkowników od punktu dostępowego \cite{kryszkiewicz2020stochastic}. 

Choć pierwsze podejście może mieć więcej punktów swobody, bardziej szczegółowo modelując zużycie energii, rozwiązanie pomiarowe jest częściej stosowane. Wynika to z jego prostoty, a także dużej dokładności przy ograniczonej liczbie parametrów. W przypadku modelowania modemów WiFi w \cite{kryszkiewicz2020stochastic} zaproponowano parametryzację z uwagi na przepływność (osobno dla nadawania i odbioru) oraz tłumienie propagacyjne między nadajnikiem i odbiornikiem. Ciekawym aspektem tego modelu jest też uwzględnienie różnic wynikających z architektury czy rozwiązań poszczególnych producentów nawet przy transmisji z użyciem tego samego standardu. W tym celu użyto zmiennych losowych. Podobny model zaproponowało niedawno 3GPP dla stacji bazowych 5G \cite{3gpp_38864}. Zużycie mocy dzieli się na część statyczną i dynamiczną. Część dynamiczna jest skalowana przez procent zajętych zasobów częstotliwościowych, liczbę anten, a także alokowaną moc. 

W przypadku przewodowej sieci rdzeniowej, służącej do przesłania zgłoszeń między węzłami MEC oraz chmurą podobny liniowy model zużycia energii jest proponowany 
\cite{idzikowski2015survey} tzn. moc stała powiększona o moc zmienną proporcjonalną do obciążenia konkretnego urządzenia sieciowego.

Warto podkreślić, że zarówno w przypadku sieci przewodowej jak i bezprzewodowej zużycie mocy konkretnego urządzenia zależy często od innych zgłoszeń/użytkowników obsługiwanych równolegle, zwiększających obciążenie konkretnych urządzeń. Zakładając, że liczba obsługiwanych urządzeń np. przez konkretny router, jest znacząca należy przyjąć pewne średnie obciążenie i dla tego punktu pracy wyznaczyć wpływ obsługi danego zgłoszenia na zużycie energii. 

Na podstawie \cite{Kopras_sensors_2024} 
można przyjąć, że całkowita energia potrzebna na transmisję zgłoszenia o długości $L$ bitów i przesłania odpowiedzi o długości $oL$ bitów wynosi 
\begin{equation}
    P_{\mathrm{comm}}=L\left(1+o\right)\left(\gamma_{w} +\gamma_{wl}\right)
\end{equation}
gdzie $\gamma_w$ oraz $\gamma_{wl}$ opisuje koszt w Joulach na bit transmisji z użyciem wybranego łącza przewodowego oraz bezprzewodowego. Warto podkreślić, że dla każdego punktu początkowego i końcowego jest możliwych zazwyczaj wiele dróg o różnym koszcie energetycznym transmisji. Analiza wybranych modeli zużycia energii pozwoliła ustalić w \cite{Bogucka_COMMAG_2023} koszt energetyczny transmisji w sieci WiFi ok. 4e4 pJ/b. Ponad trzy rzędy wielkości wyższe zużycie energii może być oczekiwane od stacji bazowej w sieci LTE. Warto porównać te dwie wartości z energią wymaganą do transmisji jednego bitu wynikająca z Twierdzenia Shannona. Wynosi ona ok. 0.55 pJ. Widać zatem, że istniejące rozwiązania radiowe są jeszcze dość odległe od granicznej efektywności energetycznej. W tym samym artykule pokazano też średnie zużycie energii na bit przez routery przewodowe. W zależności od rodzaju urządzenia to wartość ok. 20-1000 pJ.  
\subsubsection{Opóźnienie wynikające z obliczeń}
Używając oznaczeń jak w Rozdziale \ref{sec_model_energia_obliczanie} można zdefiniować opóźnienie wynikające z obliczeń jako\cite{Kopras_TCOMM_2022}: 
\begin{equation}
    D_{\mathrm{cp}}=\frac{L\theta}{fs}
\end{equation}
gdzie licznik oznacza całkowitą liczbę operacji zmiennoprzecinkowych wymaganych dla danego zadania obliczeniowego, a mianownik definiuje liczbę operacji zmiennoprzecinkowych wykonywanych w ciągu sekundy przez jednostkę obliczeniową, wynikająca z częstotliwości pracy zegara $f$ oraz liczby komend wykonywanych w jednym cyklu pracy $s$. Warto podkreślić, że jest to opóźnienie wynikające z samych obliczeń. Ważne też jest dodanie opóźnienia wynikającego z oczekiwania aż poprzednie zadanie obliczeniowe zostanie rozwiązane w danym węźle. Tutaj też ujawni się największa różnica między MEC i obliczeniami chmurowymi. Bardzo duża liczba węzłów w chmurze obliczeniowej pozwala założyć, że zgłoszenia obliczeniowe będą mogły być przypisane do wolnego węzła prawie natychmiast. w przypadku sieci MEC można przypuszczać, że bliskie, ale ograniczone węzły obliczeniowe będą wymuszały oczekiwanie w kolejce kolejnych zadań.     

\subsubsection{Opóźnienie wynikające z komunikacji}
Opóźnienie komunikacyjne możemy podzielić na 3 addytywne składowe: opóźnienie wynikające z ograniczonej przepływności łącza, opóźnienie wynikające z odległości między źródłem i ujściem wiadomości oraz opóźnienie wynikające z ograniczeń w dostępie do kanału. 

Pierwsze z nich jest wspólne zarówno dla części przewodowej i bezprzewodowej. Dla zadania obliczeniowego o długości $L$ bitów oraz odpowiedzi o długości $oL$ bitów i szybkości bitowej łącza $r$ opóźnienie można wyliczyć jako
\begin{equation}
    D_{\mathrm{comm}}=\frac{L(1+o)}{r}.
\end{equation}

Opóźnienie wynikające z odległości będzie pomijalne przy lokalnych obliczeniach z wykorzystaniem lokalnego węzła MEC. Natomiast w przypadku połączenia z chmurą obliczeniową, odległą o tysiące kilometrów, może być to istotny czynnik. Jak pokazano w \cite{Kopras_TCOMM_2022} to opóźnienie wynosi dla sieci optycznej ok. $7.5\mu s/km$. Poprzez przemnożenie tego współczynnika przez odległość uzyskujemy ten składnik opóźnienia.

Ostatnie z rozważanych źródeł opóźnień jest najbardziej znaczące w przypadku transmisji bezprzewodowej. W wyniku losowego dostępu do kanału (np. WiFi) albo konieczność retransmisji pakietów w następstwie interferencji od innych użytkowników (np. 5G) dane zgłoszenie może dotrzeć do pierwszego węzła sieci MEC opóźnione w sposób losowy. Nie jest to jednak problem z perspektywy problemu przydziału zadań do węzłów obliczeniowych. To opóźnienie już miało miejsce, więc może zostać zmierzone. Musi być jednak uwzględnione. Większym problemem jest opóźnienie w dostępie do łącza przy transmisji odpowiedzi. W artykule \cite{Kopras_TCOMM_2022} założono wykorzystanie sieci WiFi. Używając istniejących modeli matematycznych wyznaczono rozkład opóźnień w dostępie do kanału. Wykorzystano percentyl 98\% opóźnienia podczas optymalizacji. Zatem w 98\% przypadków powrót odpowiedzi powinien nastąpić przed planowaną chwilą czasu.  

\section{Zintegrowana optymalizacja efektywności energetycznej transmisji i obliczeń}
Dla tak zdefiniowanej sieci można rozważyć optymalizację alokacji zadania obliczeniowego do odpowiedniego węzła obliczeniowego (w warstwie MEC albo w chmurze), a także zegaru procesora używanego w węźle obliczeniowym. Co istotne, założono, że liczba węzłów MEC jest ściśle ograniczona podczas gdy chmur umożliwia obsługę każdego zadania obliczeniowego na niezależnym węźle. Celem minimalizacja sumarycznej energii wymaganej na obsługę zadań przy założeniu, że każde zgłoszenie ma wyznaczony maksymalny sumaryczny czas obsługi.
Jak pokazano w \cite{Kopras_TCOMM_2022} tak zdefiniowany problem należy do klasy MINLP (ang. Mixed Integer Nonlinear Programming) z uwagi na występowanie zarówno zmiennych całkowitych tzn. przypisanie zadań do węzłów, a także ciągłych tzn. częstotliwości procesorów. Dodatkowo funkcja opisująca zużycie energii przez procesor może być niewypukła. Problem rozwiązano metodą SCA (ang. Successive Convex Approximation) z perspektywy częstotliwości procesora. Użyto również algorytmu węgierskiego dla przypisania zadań do węzłów. 

\begin{figure}[H]
\centering
\includegraphics[width=3.0in]{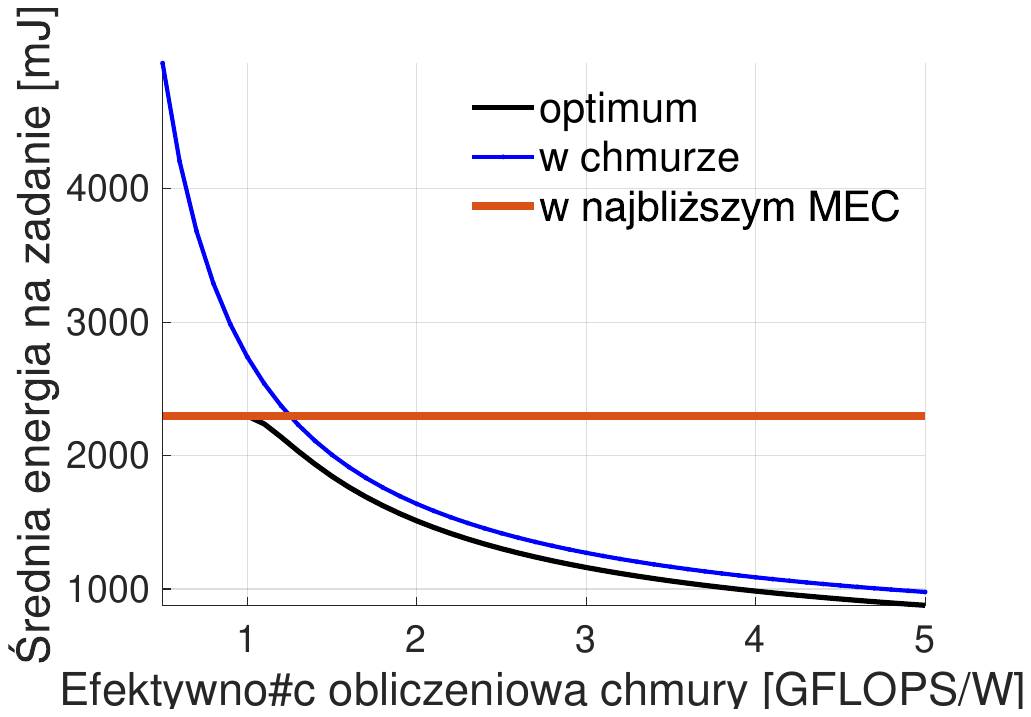}
\caption{Średnia energia na zadanie obliczeniowe w funkcji efektywności obliczeniowej chmury.}
\label{fig:srednia_energia}
\end{figure}

\begin{figure}[H]
\centering
\includegraphics[width=3.0in]{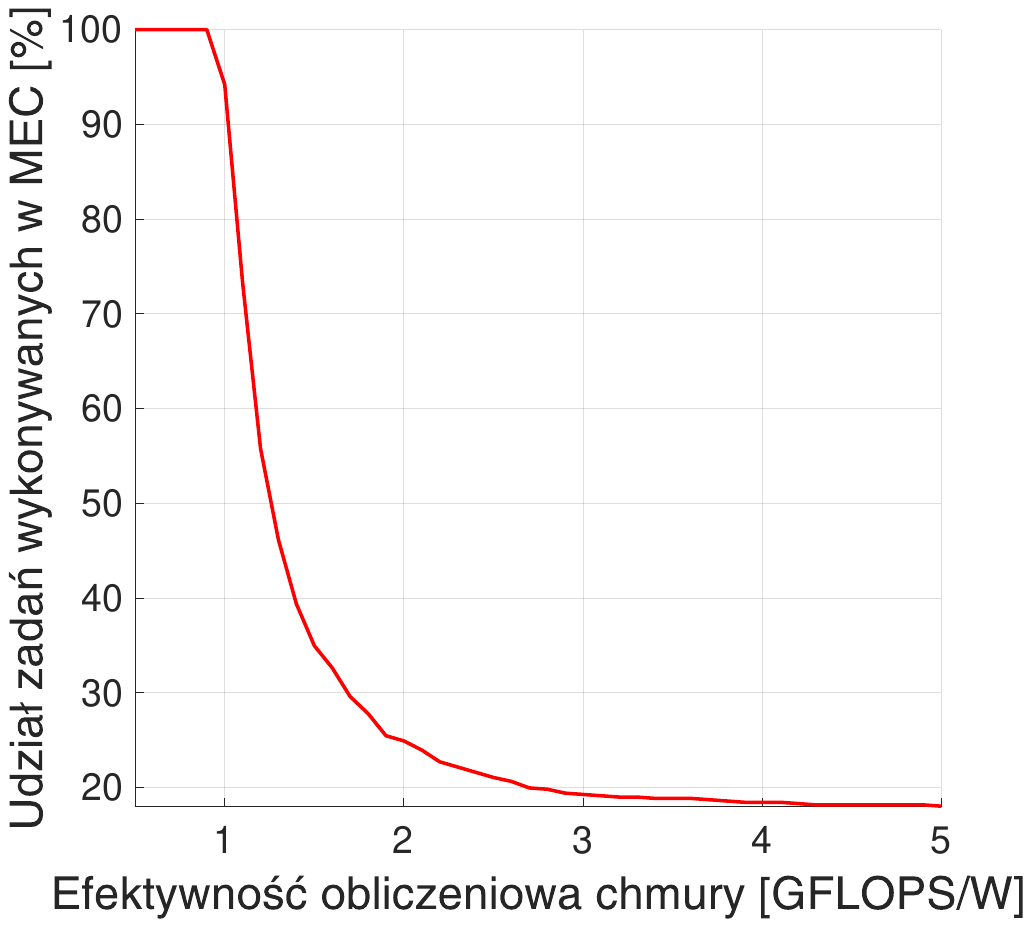}
\caption{ Procentowy udział MEC w rozwiązywaniu zadań w funkcji efektywności obliczeniowej chmury.}
\label{fig:miejsce_obliczen}
\end{figure}

\begin{figure}[H]
\centering
\includegraphics[width=3.0in]{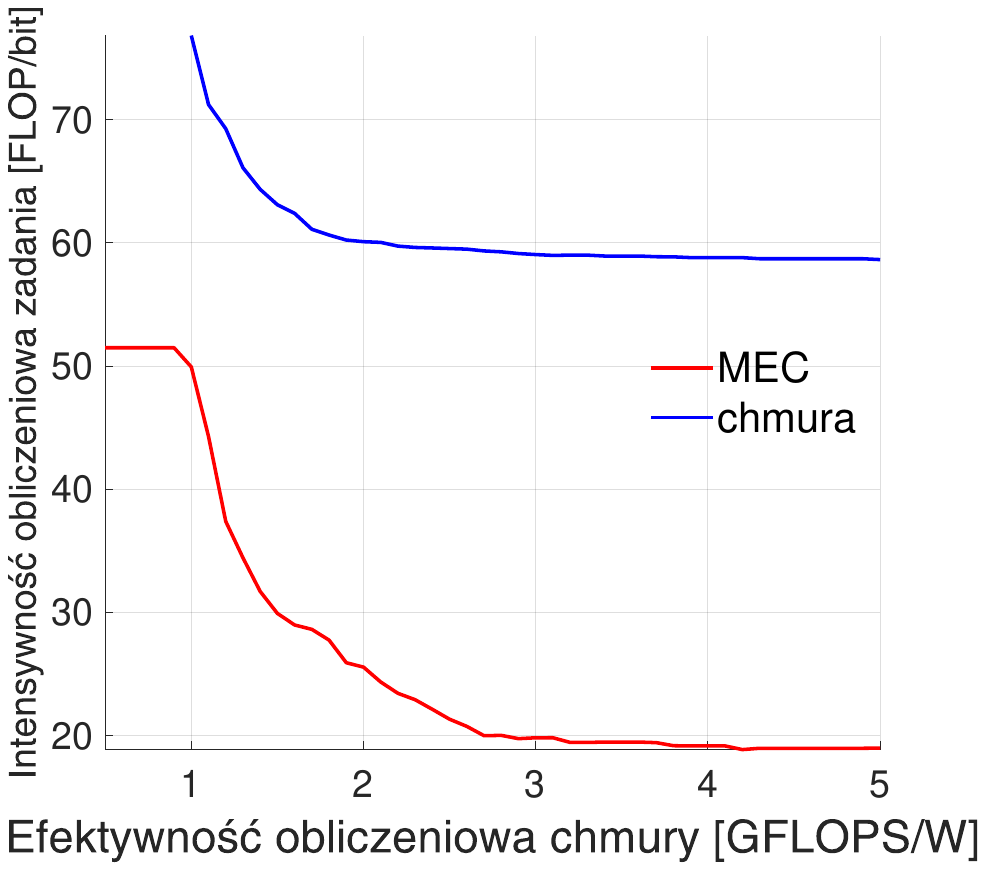}
\caption{ Średnia efektywność obliczeniowa zadań wykonywanych w chmurze i w węzłach MEC w funkcji efektywności obliczeniowej chmury.}
\label{fig:zlozonosc_a_miejsce}
\end{figure}

Parametry symulacyjne opisano w \cite{Kopras_TCOMM_2022}. Generowane są zgłoszenia o losowej długości, wymaganej intensywności obliczeniowej, a także ograniczeniu na maksymalny czas obsługi. Na Rys. \ref{fig:srednia_energia} porównano średnią energię zużytą na jedno zadanie obliczeniowe w funkcji efektywności obliczeniowej chmury dla trzech algorytmów przydziału zasobów. W algorytmie przydzielającym zadania tylko w chmurze efektywność obliczeniowa chmury ma znaczący wpływ na koszt energetyczny zadań. Widoczna jest odwrotnie proporcjonalna zależność. Jak można przypuszczać na algorytm przydzielający zadania obliczeniowe najbliższemu węzłowi MEC nie ma wpływu zmiana efektywności chmury. Najistotniejsze jest jednak, że zaproponowane rozwiązanie optymalne, dzięki rozdzieleniu zadań między wiele węzłów MEC oraz chmurę może osiągać najniższe średnie zużycie energii na zadanie. Efektywność obliczeniowa chmury ma wpływ na ten przebieg. Co istotne, nawet dla bardzo wydajnej obliczeniowo chmury widoczny jest pewien zysk poprzez wykorzystanie zasobów MEC. Żeby przeanalizować to zjawisko można spojrzeć na Rys. \ref{fig:miejsce_obliczen} gdzie pokazane jest procent zadań obliczeniowych wykonywanych w węzłach MEC przy optymalnej alokacji. Choć, zgodnie z oczekiwaniami, 100\% zadań obliczeniowych jest wykonywanych w węzłach MEC przy niskiej wydajności obliczeniowej chmury ciekawe jest, że przy bardzo wydajnej chmurze ok. 20\% zadań obliczeniowych wykonywane jest w węzłach MEC.
Wynika to z części bardzo prostych obliczeniowo zadań, których koszt energetyczny przesłania do chmury może być znaczący. Poparciem tej hipotezy może być Rys. \ref{fig:zlozonosc_a_miejsce}, który pokazuje średnią intensywność obliczeniową $\theta$ zadań  wykonywanych w chmurze i w warstwie MEC. Widoczne jest, że średnio chmura wykorzystywana jest do bardziej złożonych zadań. Nawet przy wyższych efektywnościach obliczeniowych chmury proste zadania (o średniej intensywności obliczeniowej ok. 20 FLOP/bit) są wykonywane w warstwie MEC. Warto tu podkreślić, że powyższa analiza nie rozważa problemu odrzucania zgłoszeń niespełniających wymogów opóźnieniowych. Wpływ tego zjawiska jak również wielu możliwych parametrów przedstawiono w \cite{Kopras_TCOMM_2022}. 

\section{Nowe perspektywy optymalizacji efektywności energetycznej sieci MEC }
Jak pokazał poprzedni rozdział optymalny rozdział zadań między węzły obliczeniowe wraz z uwzględnieniem optymalizacji pracy procesorów pozwala znacząco obniżyć zużycie energii w stosunku do prostszych rozwiązań. 
W tym rozdziale zostaną przedyskutowane dwa nowe podejścia do optymalizacji zużycia mocy w sieci MEC uwzględniające dodatkowe, rzadko rozważane zmienne optymalizacyjne. 

Z perspektywy urządzenia końcowego, często zasilanego bateryjnie, istotne jest obniżenie zużycia energii na przesył zadania obliczeniowego do stacji bazowej lub punktu dostępowego. Choć można w tym celu wykorzystać ogólne metody z perspektywy komunikacyjnej \cite{Nossek_2011}, bardziej odpowiednie jest zdefiniowanie tego problemu uwzględniając potencjał obliczeniowy urządzenia końcowego \cite{Kryszkiewicz_PIMRC_2019}. W przypadku transmisji danych na duże odległości np. z odległych kamer czy czujników IoT, największy udział w zużyciu mocy urządzenia końcowego ma wzmacniacz mocy. Można stosować różne architektury wzmacniaczy, jak również przetwarzanie wstępne sygnału, żeby zwiększyć efektywność energetyczną transmisji \cite{Kryszkiewicz_Sroka_Hoffmann_Wachowiak_2023}. Stosunkowo prostym rozwiązaniem jest dobór punktu pracy wzmacniacza, tak, żeby zmaksymalizować jakość sygnału w punkcie odbioru albo efektywność energetyczną transmisji. Istotnym jest, że taki dobór punktu pracy powoduje pojawienie się zniekształceń nieliniowych w pasmie pracy nadajnika, ale przy zwiększeniu mocy sygnału pożądanego docierającego do odbiornika albo zmniejszeniu zużycia energii na transmisję. 
Zazwyczaj standardy bezprzewodowe ograniczają poziom zniekształceń po stronie nadajnika np. przez ustalony poziom EVM (ang. Error Vector Magnitude), nie jest to jednak optymalne z perspektywy odbiornika. 

\begin{figure}[H]
\centering
\includegraphics[width=3.0in]{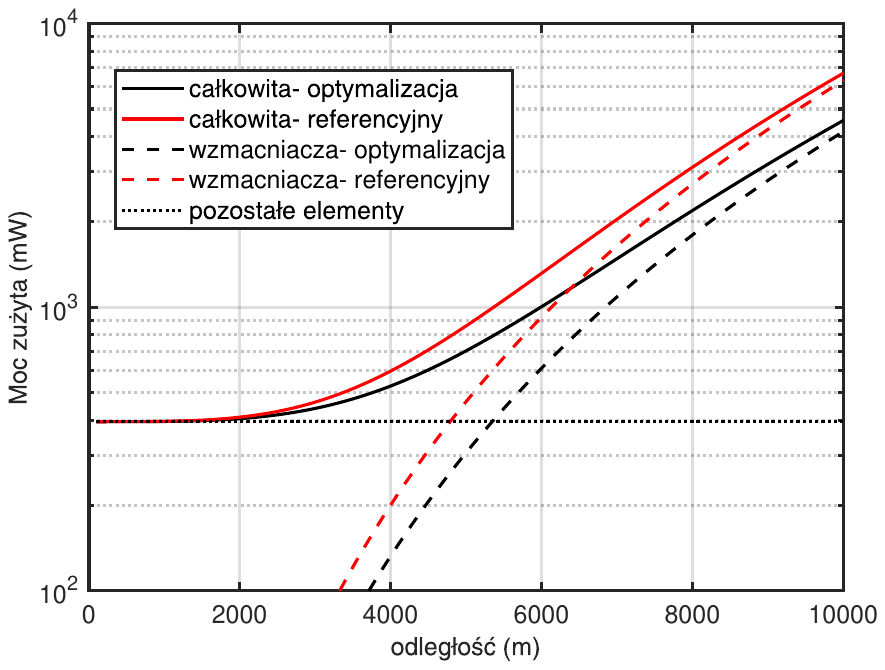}
\caption{Moc zużywana przez urządzenie końcowe na przesył zadań obliczeniowych: scenariusz referencyjny i z optymalizacją punktu pracy wzmacniacza}
\label{fig:UE_moc}
\end{figure}

Stosując metodologię przedstawioną w \cite{Kryszkiewicz_PIMRC_2019} wyliczono moc zużywaną na wysłanie zgłoszenia obliczeniowego (strumienia wideo o przepływności 1 Mbps) do punktu dostępowego. Założono, że transmisja odbywa się na częstotliwości 3.5GHz z użyciem technologii LTE używającej kanału 20MHz. Poza wzmacniaczem mocy uwzględniono moc zużywaną na kodowanie źródłowe sygnału wideo, kodowanie nadmiarowe, przetwarzanie cyfrowo-analogowe, modulację OFDM, oscylator lokalny, a także mikser. Jak widać na Rys. \ref{fig:UE_moc} moc zużywana przez te komponenty (poza wzmacniaczem mocy) to ok. 400 mW i jest niezależna od odległości. Widoczne jest, że w większej odległości między nadajnikiem i odbiornikiem moc zużywana przez wzmacniacz jest dominującym składnikiem. Widać też że moc wzmacniacza z optymalizacją punktu pracy jest niższa niż w przypadku systemu referencyjnego, który używa stałego punktu pracy określonej przez IBO (ang. Input Back-Off) na 6 dB. Optymalizacja pozwala ograniczyć całkowitą moc zużywaną przez nadajnik o ok. 32 \% w odległości 10 km.

Ostatnim stopniem swobody, który warto rozważyć przy optymalizacji efektywności energetycznej sieci MEC jest wiek informacji AoI (ang. Age of Information) \cite{Bogucka_COMMAG_2023}. Często zgłoszenia obliczeniowe wysyłane są do sieci MEC okresowo z danego czujnika. Zwiększenie częstotliwości wysyłania takich danych powoduje zwiększenie wymaganej mocy obliczeniowej, a także zwiększa obciążenie węzłów sieciowych. W sytuacji ekstremalnej może to spowodować skolejkowanie kolejnych zgłoszeń zwiększając dodatkowo wiek informacji w momencie przetwarzania. Jednocześnie często kolejne dane pomiarowe mogą być silnie skorelowane, a przez to nadmiarowe. Z tej perspektywy sensowne wydaje się minimalizacja częstotliwości przesyłania danych z sensora tak, żeby zapewnić wymagane wskaźniki jakości danej aplikacji (np. wymagane prawdopodobieństwo wykrycia zdarzeń niebezpiecznych przy monitoringu wizyjnym). Wymaga to jednak zbudowania modelu pętli sterowania dla danego typu zastosowania i uwzględnienie jej w globalnym zadaniu  optymalizacyjnym sieci MEC. Przykładem może być optymalizacja komunikacji w konwoju pojazdów pod kątem ograniczenia prawdopodobieństwa wystąpienia kolizji między pojazdami\cite{Sroka_sensors_2023}.

 \section{Podsumowanie}
W artykule pokazano, że wykorzystanie systemu MEC może pozwolić zapewnić wiele nowych usług szczególnie wymagających niskich opóźnień. Z perspektywy ograniczenia zużycia energii oraz zapewnienia wysokiej jakości usług MEC konieczny jest wyspecjalizowany przydział zadań obliczeniowych. Wymaga to jednak znajomości modelu całej sieci komunikacyjnej i obliczeniowej, a także możliwości wpływania na konfigurację na wszystkich etapach przesyłu i przetwarzania. Choć jest to zadanie wymagające może znacząco poprawić efektywność działania takiej sieci.
 
\bibliographystyle{krit}
\bibliography{references}

\end{multicols}
\end{document}